\documentclass{ws-ijtaf}

\usepackage{graphics}
\usepackage{psfrag}
\usepackage{amsmath}
\usepackage{amssymb}

\begin{document}

\markboth{Lopez-Ruiz, Shivanian, Abbasbandy, Lopez}
{A Generalized Continuous Model for Random Markets}

\catchline{}{}{}{}{}
%%%%%%%%%%%%%%%%%%%%%%%%%%%%%%%%%%%%%%%%%%%%%%%%%%%%%%%%%%%%%%%%%%%%

\title{A GENERALIZED CONTINUOUS MODEL \\ FOR RANDOM MARKETS}

\author{RICARDO LOPEZ-RUIZ}
\address{DIIS \& BIFI, University of Zaragoza, \\
Zaragoza, 50009, Spain \\
\email{rilopez@unizar.es}}

\author{ELYAS SHIVANIAN}
\address{Dept. of Mathematics, Imam Khomeini International University, \\
Qazvin, 34149-16818, Iran \\
\email{shivanian@ikiu.ac.ir}}

\author{SAIED ABBASBANDY}
\address{Dept. of Mathematics, Imam Khomeini International University, \\
Qazvin, 34149-16818, Iran \\
\email{abbasbandy@yahoo.com}}

\author{JOSE-LUIS LOPEZ}
\address{Dept. of Math. Engineering, Public University of Navarre, \\
Pamplona, 31006, Spain \\
\email{jl.lopez@unavarra.es}}

\maketitle

\begin{history}
%\received{(Day Month Year)}
%\revised{(Day Month Year)}
%\accepted{(Day Month Year)}
%\comby{(xxxxxxxxxx)}
\end{history}

\begin{abstract}
A generalized continuous economic model is proposed for random markets.
In this model, agents interact by pairs and exchange their money in
a random way. A parameter controls the effectiveness of the transactions
between the agents. We show in a rigorous way that this type of markets
reach their asymptotic equilibrium on the exponential wealth distribution.
\end{abstract}



\section{Introduction}
\label{intro}

In the last years, different techniques and models from statistical physics are being
successfully applied to explain some real data observed in economy \cite{chakra2010}.
Concretely, it has been reported that western (capitalistic) economies can be divided into
two distinct groups according to the incomes of people \cite{dragu2001}.
The 95\% of the population, the middle and lower economic classes of society,
arranges their incomes in an exponential wealth distribution.
The incomes of the rest of the population, the 5\% of individuals,
fit a power law distribution.

Different models have been used to explain the origin of these wealth distributions.
From a macroscopic point of view, it is necessary to take into account that
markets have an intrinsic random ingredient as a consequence of the interaction
of an undetermined ensemble of agents which are performing an undetermined number
of commercial transactions at each moment. A kind of models considering this unknowledge
associated to markets are the gas-like models \cite{yakoven2009}.
These random models interpret economic exchanges of money between agents
similarly to collisions in a gas where particles share their energy.
In order to explain the two different types of statistical behavior before mentioned,
different gas-like models have been proposed.
On the one hand, the exponential distribution can be obtained by supposing a gas
of agents that trade with money in binary collisions, or in first-neighbor interactions,
and where the agents are selected in a random, deterministic or chaotic way
\cite{dragu2000,estevez2008,pellicer2011}. On the other hand, the power law
behavior can be simulated by introducing inhomogeneity in the gas by different methods,
through the breaking of the pairing symmetry in the exchange rules \cite{pellicer2011}
or through the saving propensity of the agents \cite{chatter2004}.

In this work, we consider a continuous version of one of such
homogeneous gas-like models \cite{lopezruiz2011} which we generalize to a situation
where the agents present a control parameter to decide the degree of interaction with
the rest of economic agents. A zero value of the parameter will represent a null interaction
of each agent with the rest of the ensemble and a value equal to unity will mean
the total disponibility of each agent for the interaction with the whole gas.
Under these circumstances, we are interested in explaining the appearance
of the exponential (Gibbs) distribution for all the intermediary values,
between zero and one, of the parameter value. This situation can represent the ubiquity
of this distribution in many natural phenomena but in particular in the random markets.
Thus, the rigorous analytical results here exposed will help us to enlighten the behavior
of real data for income distributions but also to understand the computational statistical
behavior found in such type of economic models which decay to the
exponential wealth distribution in their asymptotic regime
independently of the initial wealth distribution given to the system.

\section{The continuous gas-like model. Properties}
\label{sec:1}

We consider an ensemble of economic agents trading with money in a random manner \cite{dragu2000}.
This is one of the simplest gas-like models, in which an initial amount of money is given
to each agent, let us suppose the same to each one. Then, pairs of agents are randomly chosen
and they exchange their money also in a random way. When the gas evolves under these conditions,
the exponential distribution appears as the asymptotic wealth distribution. In this model,
the microdynamics is conservative because the local interactions conserve
the money. Hence, the macrodynamics is  also conservative and the total amount of money is
constant in time.

The discrete version of this model is as follows \cite{dragu2000}.
The trading rules for each interacting pair $(m_i,m_j)$ of the ensemble of $N$ economic
agents can be written as
\begin{eqnarray}
m'_i &=& \epsilon \; (m_i + m_j), \nonumber\\
m'_j &=& (1 - \epsilon) (m_i + m_j), \label{model1}\\
i , j &=& 1 \ldots N, \nonumber
\end{eqnarray}
where $\epsilon$ is a random number in the interval $(0,1)$.
The agents $(i,j)$ are randomly chosen. Their initial money $(m_i, m_j)$,
at time $t$, is transformed after the interaction in $(m'_i, m'_j)$ at time $t+1$.
The asymptotic distribution $p_f(m)$, obtained by numerical simulations,
is the exponential (Boltzmann-Gibbs) distribution,
\begin{equation}
p_f(m)=\beta \exp(-\beta \,m), \hspace{0.5cm}\hbox{with}\hspace{0.5cm}\beta={1/ <m>_{gas}},
\label{eq-exp}
\end{equation}
where $p_f(m) {\mathrm d}m$ denotes the PDF ({\it probability density function}), i.e.
the probability of finding an agent with money (or energy in a gas system) between
$m$ and $m + {\mathrm d}m$.
Evidently, this PDF is normalized, $\vert\vert p_f\vert\vert=\int_0^{\infty} p_f(m){\mathrm d}m=1$.
The mean value of the wealth, $<m>_{gas}$, can be easily calculated directly from the gas
by $<m>_{gas}=\sum_i m_i/N$.

The continuous version of this model \cite{lopezruiz2011} considers the evolution of
an initial wealth distribution $p_0(m)$ at each time step $n$ under the action of an operator $T$.
Thus, the system evolves from time $n$ to time $n+1$ to asymptotically
reach the equilibrium distribution $p_f(m)$, i.e.
\begin{equation}
\lim_{n\rightarrow\infty} {T}^n \left(p_0(m)\right) \rightarrow p_f(m).
\label{eq-operT}
\end{equation}
In this particular case, $p_f(m)$ is the exponential distribution with the same
average value $<p_f>$ than the initial one $<p_0>$,
due to the local and total richness conservation.

The derivation of the operator $T$ is as follows \cite{lopezruiz2011}.
Suppose that $p_n$ is the wealth distribution in the ensemble at time $n$.
The probability to have a quantity of money $x$ at time $n+1$ will be the sum of the
probabilities of all those pairs of agents $(u,v)$ able to
produce the quantity $x$ after their interaction, that is, all the pairs verifying $u+v>x$.
Thus, the probability that two of these agents with money $(u,v)$ interact between them is
$p_n(u)*p_n(v)$. Their exchange is totally random and then they can give rise with equal
probability to any value $x$ comprised in the interval $(0,u+v)$. Therefore, the probability
to obtain a particular $x$ (with $x<u+v$) for the interacting pair $(u,v)$ will be
$p_n(u)*p_n(v)/(u+v)$.  Then, $T$ has the form of a nonlinear integral operator,
\begin{equation}
p_{n+1}(x)={T}p_n(x) = \int\!\!\int_{u+v>x}\,{p_n(u)p_n(v)\over u+v}
\; {\mathrm d}u{\mathrm d}v \,.
\label{eq-T}
\end{equation}

If we suppose $T$ acting in the PDFs space, it has been proved \cite{lopez2011}
that $T$ conserves the mean wealth of the system, $<Tp>=<p>$. It also conserves
the norm ($\vert\vert \cdot\vert\vert$), i.e. $T$ maintains the total number of agents
of the system, $\vert\vert T p\vert\vert=\vert\vert p\vert\vert=1$, that
by extension implies the conservation of the total richness of the system.
We have also shown that the exponential distribution $p_f(x)$ with the right average value
is the only steady state of $T$, i.e. $T p_f=p_f$. Computations also seem to suggest
that other high period orbits do not exist.
In consequence, it can be argued that the relation (\ref{eq-operT}) is true.
We proceed to recall these properties and to establish some new ones.

\subsection{Properties of the operator $T$. Recall}

First, in order to set up the adequate mathematical framework,
we provide the following definitions.

\begin{definition}
We introduce the space $L_1^+$ of positive functions (wealth distributions)
in the interval $[0,\infty)$,
$$
L_1^+[0,\infty)=\lbrace y:[0,\infty)\to R^+\cup\lbrace0\rbrace,
\hskip 2mm \vert\vert y\vert\vert<\infty\rbrace,
$$
with norm
$$
\vert\vert y\vert\vert=\int_0^\infty y(x) dx.
$$
\end{definition}

\begin{definition}
\label{def-mean1}
We define the mean richness $<x>_y$ associated to a wealth distribution $y\in L_1^+[0,\infty)$ as
the mean value of $x$ for the distribution $y$. In the rest of the paper,
we will represent it by $<y>$. Then,
$$
<y> \equiv <x>_y = \vert\vert xy(x)\vert\vert=\int_0^\infty xy(x) dx.
$$
\end{definition}

\begin{definition}
\label{def-region}
For $x\ge 0$ and $y\in L_1^+[0,\infty)$ the action of operator $T$ on $y$ is defined by
$$
T(y(x))=\int\int_{S(x)} dudv{y(u)y(v)\over u+v},
$$
with $S(x)$ the region of the plane representing the pairs of agents $(u,v)$ which can
generate a richness $x$ after their trading, i.e.
$$
S(x)=\lbrace (u,v), \hskip 2mm u,v>0,\hskip 2mm u+v>x\rbrace.
$$
\end{definition}

Now, we remind the following results recently presented in Ref. \cite{lopez2011}.

\begin{theorem}
\label{teorema-norma}
For any $y\in L_1^+[0,\infty)$ we have that
$\vert\vert Ty\vert\vert=\vert\vert y\vert\vert^2$.
(It means that the number of agents in the economic system is conserved in time).
In particular, consider the subset of PDFs in $L_1^+[0,\infty)$, i.e. the unit sphere
$B=\lbrace y\in L_1^+[0,\infty)$, $\vert\vert y\vert\vert=1\rbrace$. Observe that
if $y\in B$ then $Ty\in B$.
\end{theorem}

\begin{theorem}
The mean value $<y>$ of a PDF $y$ is conserved, that is $<Ty>=<y>$ for any $y\in B$.
(It means that the mean wealth, and by extension the total richness, of the economic
system are preserved in time).
\end{theorem}

\begin{theorem}
Apart from $y=0$, the one-parameter family of functions
$y_\alpha(x)= \alpha e^{-\alpha x}$, $\alpha>0$, are the unique fixed points
of $T$ in the space $L_1^+[0,\infty)$.
\label{teorema-unicidad}
\end{theorem}

\begin{proposition}
The operator $T$ is Lipschitz continuous in $B$ with Lipschitz constant $\le 2$.
It means that if we take $y,w\in B$, then
$$
\vert\vert Ty-Tw \vert\vert \le 2\,\vert\vert y-w\vert\vert.
$$
\end{proposition}

\begin{proposition} 
For any $y\in L_1^+[0,\infty)$ and $m\le n\in N$, 
it holds that $(-1)^m(T^ny)^{(m)}\in L_1^+[0,\infty)$.
It implies that if $T^ny=y$ then $y$ is a completely monotonic function.
\label{propo-CMF}
\end{proposition}

\subsection{Other properties of $T$}

Evidently, our main interest resides in the action of the operator $T$
in the subset of PDFs. We give here other properties for this
restriction of $T$ in $B$, although most of these properties are clearly
valid for the whole space $L_1^+[0,\infty)$.

\begin{proposition}
Suppose that $y\in B$ then $Ty$ is a decreasing function.
\label{propo-decreasing}
\end{proposition}

\begin{proof}
Assume that $x_1,x_2\ge 0$ and $x_1\le x_2$ then
$\lbrace (u,v), \hskip 2mm u,v>0,\hskip 2mm u+v>x_2\rbrace
\subset\lbrace (u,v), \hskip 2mm u,v>0,\hskip 2mm u+v>x_1\rbrace$ and so
$$
Ty(x_1) = \int\!\!\int_{u+v>x_1}\,{y(u)y(v)\over u+v}\;{\mathrm d}u{\mathrm d}v \,\ge
\int\!\!\int_{u+v>x_2}\,{y(u)y(v)\over u+v}\;{\mathrm d}u{\mathrm d}v = Ty(x_2)\,.
$$
Therefore $Ty$ for every function $y\in B$ is a decreasing function.
This is a particular case of Proposition \ref{propo-CMF} for $n=m=1$.
\end{proof}

\begin{corollary}
Suppose that $y\in B$ is a not decreasing function then
$\forall n\in\mathbb{N}: T^ny\neq y$. 
\end{corollary}

\begin{proof}
It follows directly from Proposition \ref{propo-decreasing}.
\end{proof}

\begin{proposition}
\begin{arabiclist}
 \item For every $y,w\in B$ we have $\vert\vert Ty-Tw\vert\vert\leq 2$.
 \item For some members $y,w\in B$,
        $\vert\vert Ty-Tw\vert\vert=\vert\vert y-w\vert\vert$, hence $T$ is not a contraction.
\end{arabiclist}
\end{proposition}

\begin{proof}
\begin{arabiclist}
 \item It is clear that $\vert\vert Ty-Tw\vert\vert\leq \vert\vert Ty\vert\vert +
                         \vert\vert Tw\vert\vert = 1+1 = 2$.
 \item Take $y=\alpha e^{-\alpha x}$ and $w=\beta e^{-\beta x}$, with $\alpha,\beta>0$,
 which belong to $B$. In this case, we can easily see that $\vert\vert Ty-Tw\vert\vert=
 \vert\vert\alpha e^{-\alpha x}-\beta e^{-\beta x}\vert\vert=\vert\vert y-w\vert\vert$.
\end{arabiclist}
\end{proof}

\begin{example}
Take $y(x)={1\over(1+x)^2}$ and $w(x)=e^{-x}$ which belong to $B$. By using Mathematica,
it is seen that $\vert\vert y-w\vert\vert=0.407264$ and $\vert\vert Ty-Tw\vert\vert=0.505669$.
Then, $\vert\vert Ty-Tw\vert\vert>\vert\vert y-w\vert\vert$ which confirms again that $T$ is not
a contraction in $B$.
\end{example}

\vskip 0.3cm
{\bf Conjecture:}
For any $y\in B$, we guess by simulation of many examples the truth of the following relation:
$$
\lim_{n\rightarrow\infty} {T}^ny(x)=\left\{\begin{array}{lcl}
\delta e^{-\delta x} & & with \hskip 5mm \delta=1/<y>\,,\\
& or & \\
0^+ & & when \hskip 5mm <y>=+\infty\,.
\end{array}\right.
$$

\vskip 0.3cm
Let us observe that the above pointwise limit of $T^ny$ when $n\to\infty$ can be outside of $B$
in the case that $<y>=+\infty$. See the next example.

\begin{figure}[h]
\begin{center}
\psfrag{B}{} \psfrag{A}{\large\scriptsize (a)}
\includegraphics[width=1.5in, height=1.3in]{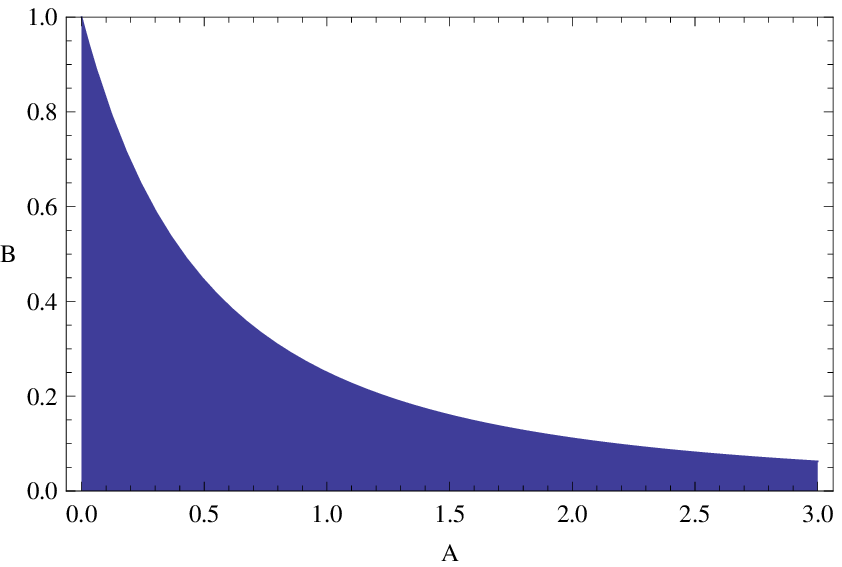} \hskip 2 mm
\psfrag{B}{} \psfrag{A}{\large\scriptsize (b)}
\includegraphics[width=1.5in, height=1.3in]{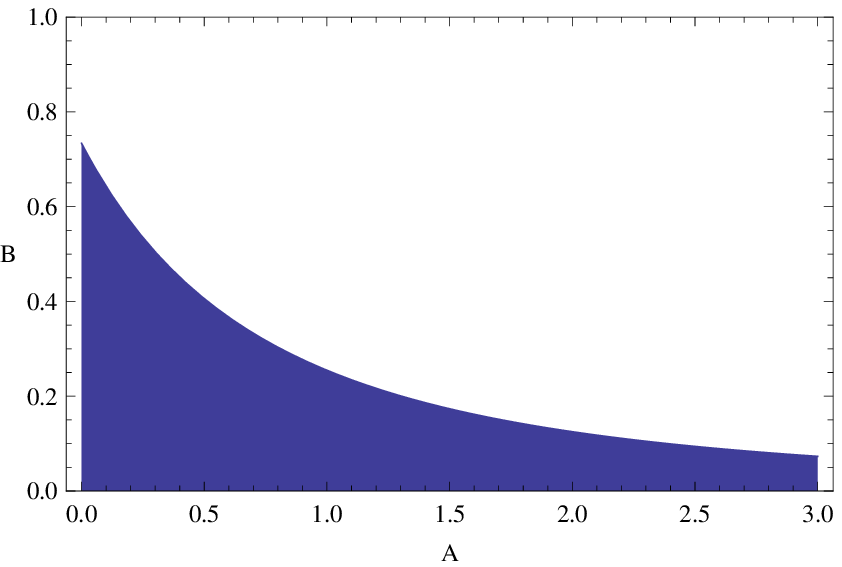} \hskip 2 mm
\psfrag{B}{} \psfrag{A}{\large\scriptsize (c)}
\includegraphics[width=1.5in, height=1.3in]{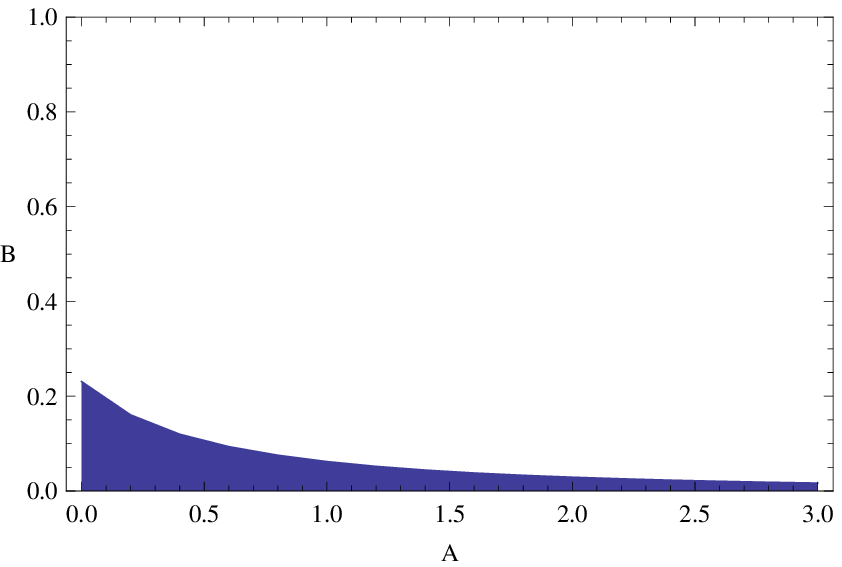}
\caption{(a) $y(x)={1\over(1+x)^2}$, (b) $Ty(x)$, (c) $T^2y(x)$.}
\label{fig-outside-B}
\end{center}
\end{figure}

\begin{example}
Take $y(x)={1\over(1+x)^2}$ which belongs to $B$, with $<y>=+\infty$.
Evidently, $T^ny\in B$ for all $n$. But it can be graphically guessed that
$\lim_{n\rightarrow\infty} {T}^ny(x)=0^+\notin B$. See Figure \ref{fig-outside-B}.
\label{example-pareto}
\end{example}

\begin{theorem}
Suppose that $\lim_{n\rightarrow\infty} \vert\vert T^ny(x)-\mu(x)\vert\vert=0$, 
and $\mu(x)$ is a continuous function, then $\mu(x)$ is the fixed point
of the operator $T$ for the initial condition $y(x)\in B$.
In other words, $\mu(x)=\delta e^{-\delta x}$ with $\delta=1/<y>$.
\label{theor-T-clave}
\end{theorem}

\begin{proof}
First we prove that for each $\epsilon>0$ there exists $M$ so that $\forall n>M$:
$\vert\vert T^ny-T\mu\vert\vert<\epsilon$. Since 
$\lim_{n\rightarrow\infty} \vert\vert T^ny(x)-\mu(x)\vert\vert=0$,
we can say that $\exists N$ such that $\forall n>N$: $\vert\vert T^{n-1}y-\mu\vert\vert<\epsilon/2$.
Now, we choose $M=N$. Then, due to the Lipschitz continuity of $T$, we have
$$
\forall n>M:\vert\vert T^ny-T\mu\vert\vert\leq 2\vert\vert T^{n-1}y-\mu\vert\vert <
2\cdot {\epsilon\over 2} = \epsilon\,.
$$
Now, by uniqueness of the limit, it implies that $T\mu=\mu$. Therefore,
for the initial condition $y(x)\in B$,
$\mu(x)$ is the fixed point of $T$, and by Theorem \ref{teorema-unicidad} it means that
$\mu(x)=\delta e^{-\delta x}$ with $\delta=1/<y>$.
\end{proof}

\begin{figure}[h]
\begin{center}
\psfrag{B}{} \psfrag{A}{\large\scriptsize (a)}
\includegraphics[width=1.5in, height=1.3in]{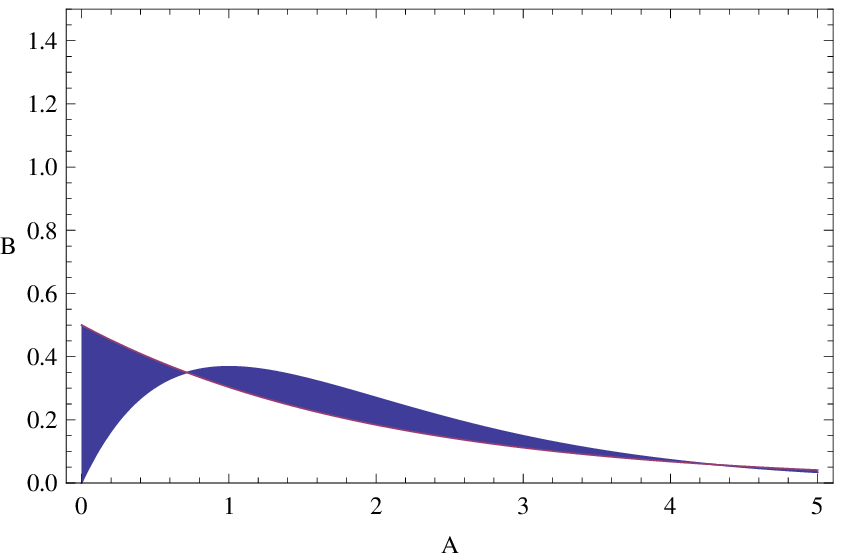} \hskip 2 mm
\psfrag{B}{} \psfrag{A}{\large\scriptsize (b)}
\includegraphics[width=1.5in, height=1.3in]{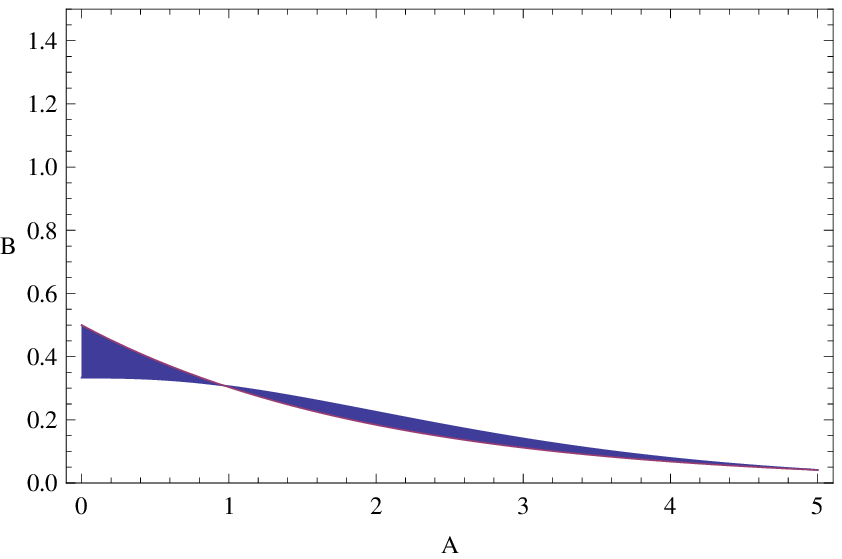} \hskip 2 mm
\psfrag{B}{} \psfrag{A}{\large\scriptsize (c)}
\includegraphics[width=1.5in, height=1.3in]{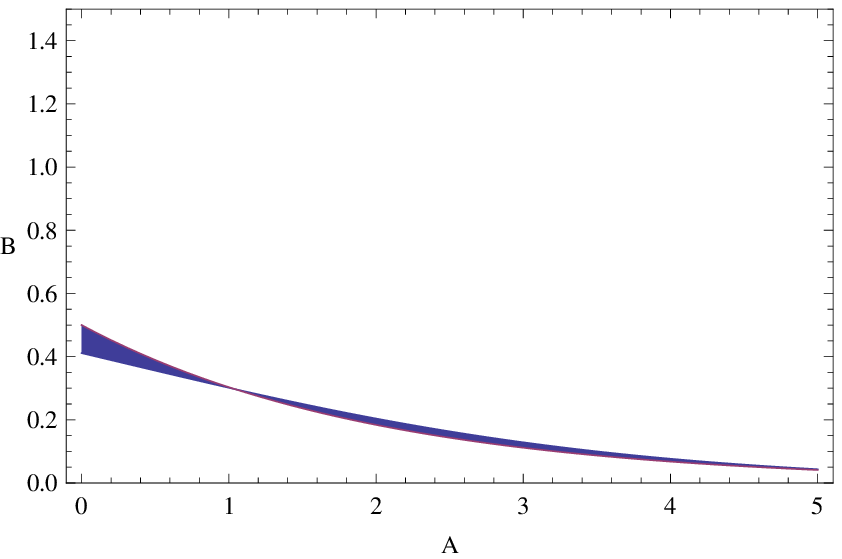}
\caption{Plot of $y(x)=xe^{-x}$, $T$-iterates of $y$ and
$\mu(x)={1\over 2}e^{-{1\over 2}x}$. (a) $\vert\vert y-\mu\vert\vert$,
(b) $\vert\vert Ty-\mu\vert\vert$, (c) $\vert\vert T^2y-\mu\vert\vert$.}
\label{fig-y-gamma}
\end{center}
\end{figure}

\begin{figure}[h]
\begin{center}
\psfrag{B}{} \psfrag{A}{\large\scriptsize (a)}
\includegraphics[width=1.5in, height=1.3in]{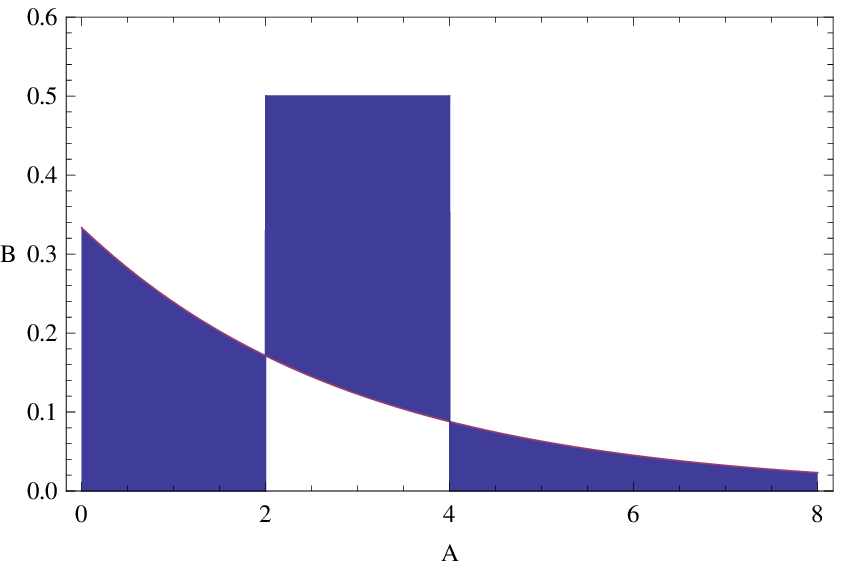} \hskip 2 mm
\psfrag{B}{} \psfrag{A}{\large\scriptsize (b)}
\includegraphics[width=1.5in, height=1.3in]{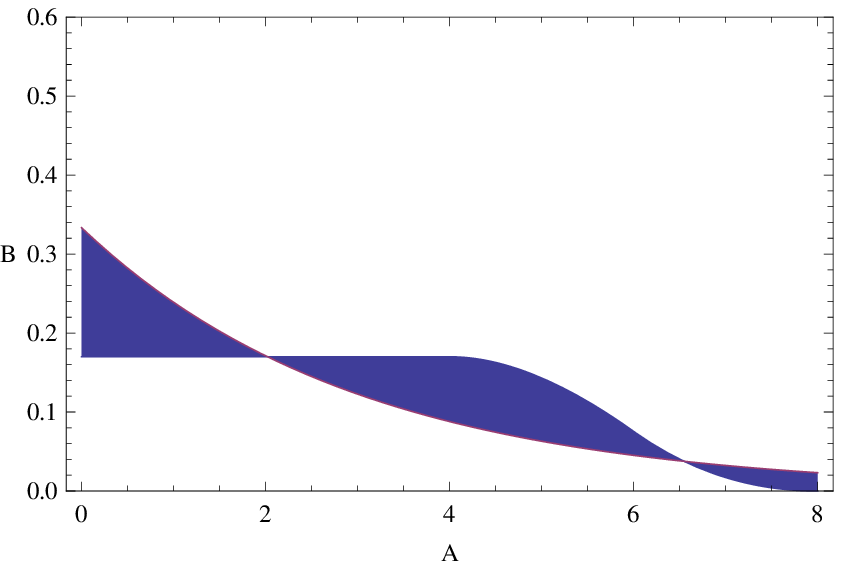} \hskip 2 mm
\psfrag{B}{} \psfrag{A}{\large\scriptsize (c)}
\includegraphics[width=1.5in, height=1.3in]{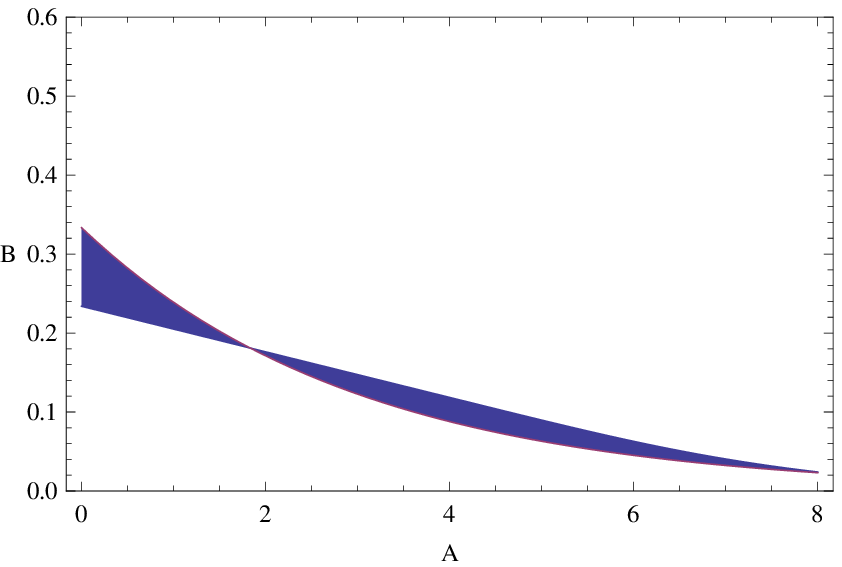}
\caption{Plot of $y(x)={1\over 2}$ if $2<x<4$, and $y(x)=0$ otherwise,
$T$-iterates of $y$ and $\mu(x)={1\over 3}e^{-{1\over 3}x}$.
(a) $\vert\vert y-\mu\vert\vert$, (b) $\vert\vert Ty-\mu\vert\vert$,
(c) $\vert\vert T^2y-\mu\vert\vert$.}
\label{fig-y-rectangular}
\end{center}
\end{figure}

As particular examples,
we present in Figs. \ref{fig-y-gamma} and \ref{fig-y-rectangular} the graphical
evidence of the convergence suggested in the former Theorem.
Many other cases have been studied with a similar behavior.
Remark that in Example \ref{example-pareto} there is pointwise convergence to
the null function ($\mu=0^+$) but there is no convergence in norm $L_1$, in fact 
$\vert\vert T^ny - \mu\vert\vert=1$ for all $n$. 

\begin{example}
Take the Gamma distribution $y(x)=xe^{-x}$, so that $y\in B$ and $\delta={1\over 2}$,
then in this case $\mu(x)={1\over 2}e^{-{1\over 2}x}$. We find numerically that
$\vert\vert y-\mu\vert\vert=0.368226$, $\vert\vert Ty-\mu\vert\vert=0.185608$,
$\vert\vert T^2y-\mu\vert\vert=0.103225$, $\vert\vert T^3y-\mu\vert\vert=0.061195$,
$\vert\vert T^4y-\mu\vert\vert=0.037675$, and so on.
It is shown in Fig. \ref{fig-y-gamma}.
Then we can guess that $\lim_{n\rightarrow\infty}\vert\vert T^ny-\mu\vert\vert=0$.
\end{example}

\begin{example}
Assume now the rectangular distribution:
$y(x)={1\over 2}$ if $2<x<4$, and $y(x)=0$ otherwise.
So, $y\in B$ and $\delta={1\over 3}$,
then in this case $\mu(x)={1\over 3}e^{-{1\over 3}x}$. We find numerically that
$\vert\vert y-\mu\vert\vert > \vert\vert Ty-\mu\vert\vert >
\vert\vert T^2y-\mu\vert\vert > \vert\vert T^3y-\mu\vert\vert$, and so on.
It is shown in Fig. \ref{fig-y-rectangular}.
Then we can also guess in this case that
$\lim_{n\rightarrow\infty}\vert\vert T^ny-\mu\vert\vert=0$.
\end{example}

Summarizing, the dynamical evolution of $T$ in $L_1^+[0,\infty)$
is roughly given by the property:
$\vert\vert Ty\vert\vert=\vert\vert y\vert\vert^2$ ,
which divides the space $L_1^+[0,\infty)$ in three regions: the interior of
the unit ball $B_{in}=\lbrace y\in L_1^+[0,\infty)$, $\vert\vert y\vert\vert<1\rbrace$,
the unit sphere $B$, and the exterior of the unit ball
$B_{ex}=\lbrace y\in L_1^+[0,\infty)$, $\vert\vert y\vert\vert>1\rbrace$.
Accordingly, the recursion
$$
y_n(x)=  T^ny_0(x), \hskip 0.5cm \hbox{with} \hskip 0.3cm y_0(x) \in  L_1^+[0,\infty),
$$
presents four different behaviors depending on $y_0$:
\begin{itemize}
\item If $y_0\in B_{in}$, then $\lim_{n\to\infty}\vert\vert y_n(x)\vert\vert=0$.

\item If $y_0\in B$ and $<y_0>$ is finite, then
$\lim_{n\to\infty}\vert\vert y_n(x)-\delta e^{-\delta x}\vert\vert=0$, with $\delta={1\over <y_0>}$.
Observe that the succession $y_n$ remains in $B$ $\forall n$.

\item If $y_0\in B$ and $<y_0>$ is infinite, then 
$\lim_{n\to\infty}\vert\vert y_n(x)-0^+\vert\vert =1$.
Observe that in this case the succession $y_n$ remains in $B$ $\forall n$ and the series $y_n$
does not converge in the norm $\vert\vert\cdot\vert\vert$ to its pointwise limit $0^+\not\in B$.

\item If $y_0\in B_{ex}$, then $\lim_{n\to\infty}\vert\vert y_n(x)\vert\vert=\infty$.
\end{itemize}

Hence, the only fixed points of the system are $y=0$ and $\delta e^{-\delta x}$.
They are asymptotically reached depending on the initial average value $<y_0>$, which
determines the final equilibrium. We proceed now to show that this behavior
is essentially maintained in the extension of this model for more general random markets.

\section{Generalized continuous model for random markets}

Let us observe that the action of operator $T$ can receive a macroscopic
interpretation respect to Eqs. (\ref{model1}) in the sense that,
each iteration of the operator $T$ means that many interactions, of order $N/2$,
have taken place between different pairs of agents. As $n$
indicates the time evolution of $T$, we can roughly assume that $t\approx N*n/2$,
where $t$ follows the microscopic evolution of the individual transactions (or collisions)
between the agents (this alternative microscopic interpretation can be seen in \cite{calbet2011}).

Let us think now that many of the economical transactions planned in markets are
not successful and they are finally frustrated. It means that markets are not totally
effective. We can reflect this fact in our models in a qualitative way by defining a
parameter $\lambda\in [0,1]$ which indicates the {\it degree of effectiveness} of the
random market. When $\lambda=1$ the market will have total effectiveness and all the
operations will be performed under the action of the random rules (\ref{model1}).
The evolution of the system in this case is given by the operator $T$.
When $\lambda=0$, all the operations
become frustrated, there is no exchange of money between the agents and then
the market stays frozen in its original state. The operator representing this type
of dynamics is just the identity operator. Therefore, we can establish a {\it generalized
continuous economic model} whose evolution in the PDFs space is determined by the
operator $T_{\lambda}$, which depends on the parameter $\lambda$ as follows:

\begin{definition}
$$
T_{\lambda}y(x)=(1-\lambda)y(x)+\lambda Ty(x), \hskip 0.5cm with \hskip 0.5cm \lambda\in[0,1].
$$
\end{definition}

Observe that the parameter $(1-\lambda)$ can also be interpreted as
a kind of {\it saving propensity} of the agents, in such a way that for $\lambda=1$
they do not save anything and they game all their resources,
and for $\lambda=0$ they save the totality of their money and
then all the transactions are frustrated and the market stays in a frozen state.

\subsection{Properties of the operator $T_{\lambda}$}

Now, we proceed to present the properties of the operator $T_{\lambda}$,
which shows a dynamical behavior essentially similar to the behavior of $T$.
Concretely, one of our main results is that the exponential distribution is also
the asymptotic wealth distribution reached by the system governed by $T_{\lambda}$,
independently of the effectiveness $\lambda$ of the random market.

Let us observe that
$T_{\lambda}=I$ for $\lambda=0$ and $T_{\lambda}=T$ for $\lambda=1$, where $I$
is the identity operator.

\begin{proposition}
$T_{\lambda}$ conserves the norm, i.e., for each $y\in B$, we have $T_{\lambda}y\in B$.
\end{proposition}

\begin{proof}
Consider $y(x)\ge 0$, then $T_{\lambda}y(x)\ge 0$. The norm is:
$$
\vert\vert T_{\lambda}y(x)\vert\vert = (1-\lambda)\vert\vert y(x)\vert\vert
+ \lambda\vert\vert Ty(x)\vert\vert = (1-\lambda)+\lambda = 1.
$$
\end{proof}

\begin{proposition}
$T_{\lambda}$ conserves the average value of $y\in B$, i.e. $<y>=<T_{\lambda}y>$,
where $<y>$ represents the mean value $<x>_y$ as expressed in Definition \ref{def-mean1}.
\end{proposition}

\begin{proof}
Take $y\in B$, then
$$
<T_{\lambda}y> = (1-\lambda)<y> + \lambda <Ty> = (1-\lambda)<y> + \lambda <y> = <y>.
$$
\end{proof}

\begin{proposition}
The operator $T_{\lambda}$ is Lipschitz continuous in $B$ with a Lipschitz factor $\alpha$
such that $\alpha\leq 1+\lambda$.
\end{proposition}

\begin{proof}
Suppose that $y,w\in B$, then
$$
\vert\vert T_{\lambda}y-T_{\lambda}w\vert\vert =
\vert\vert (1-\lambda)y+\lambda Ty-(1-\lambda)w-\lambda Tw\vert\vert \leq
$$
$$
\leq (1-\lambda)\vert\vert y-w\vert\vert + \lambda\vert\vert Ty-Tw\vert\vert \leq
$$
$$
\leq (1-\lambda)\vert\vert y-w\vert\vert + 2\lambda\vert\vert y-w\vert\vert =
$$
$$
\leq (1+\lambda)\vert\vert y-w\vert\vert.
$$
\end{proof}

\begin{theorem}
For any $\lambda\in (0,1)$, the operators $T$ and $T_{\lambda}$ have the same fixed points.
\label{theor-same}
\end{theorem}

\begin{proof}
(i) Suppose $y$ is a fixed point of $T$, i.e. $Ty=y$. Then we have $T_{\lambda}y=y$,
because $T_{\lambda}y=(1-\lambda)y+\lambda Ty=y-\lambda y+\lambda y=y$. \newline
(ii) Now suppose $y$ is a fixed point of $T_{\lambda}$, i.e. $T_{\lambda}y=y$. 
Then we have $Ty=y$, because $\lambda Ty=T_{\lambda}y-(1-\lambda)y=y-y+\lambda y=\lambda y$.
\end{proof}

\begin{corollary}
The function $y(x)=0$ and the family of exponential distributions
$y_{\delta}(x)=\delta e^{-\delta x}$, $\delta>0$, are the only fixed points
of $T_{\lambda}$ in $L_1^+[0,\infty)$, with $\lambda\in (0,1]$.
\end{corollary}

\begin{proof}
It follows from Theorems \ref{theor-same} and \ref{teorema-unicidad}.
\end{proof}

\begin{theorem}
Suppose that for a given $\lambda\in (0,1)$ we have 
$\lim_{n\to\infty}\vert\vert T_{\lambda}^ny(x)-\mu(x)\vert\vert=0$, with $\mu(x)$
a continuous function, then $\mu(x)$ should be the fixed point of the operator 
$T_{\lambda}$ for the initial condition $y(x)\in B$. In other words,
$\mu(x)=\delta e^{-\delta x}$ with $\delta={1 \over <y>}$.
\end{theorem}

\begin{proof}
Identical to the proof of Theorem \ref{theor-T-clave} by changing $T$ by $T_{\lambda}$
and taking into account Theorem \ref{theor-same}.
\end{proof}

\begin{figure}[h]
\begin{center}
\psfrag{B}{} \psfrag{A}{\large\scriptsize (a)}
\includegraphics[width=1.5in, height=1.3in]{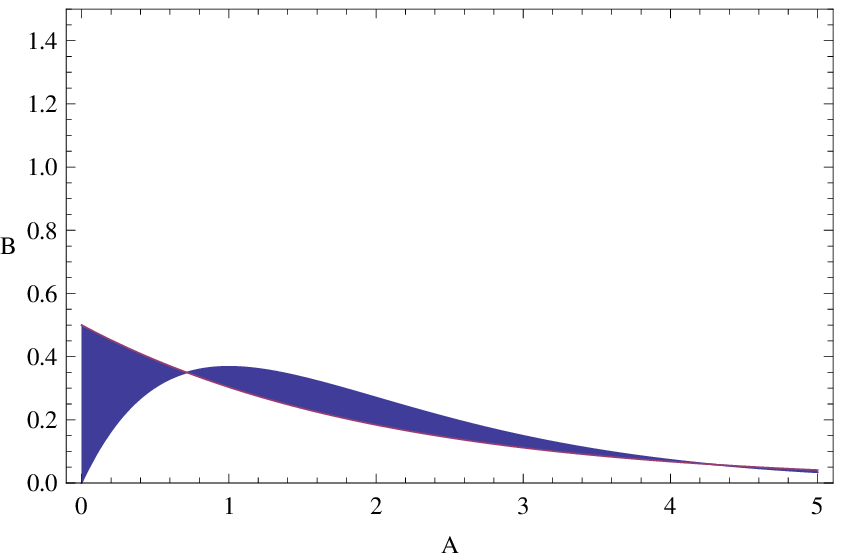} \hskip 2 mm
\psfrag{B}{} \psfrag{A}{\large\scriptsize (b)}
\includegraphics[width=1.5in, height=1.3in]{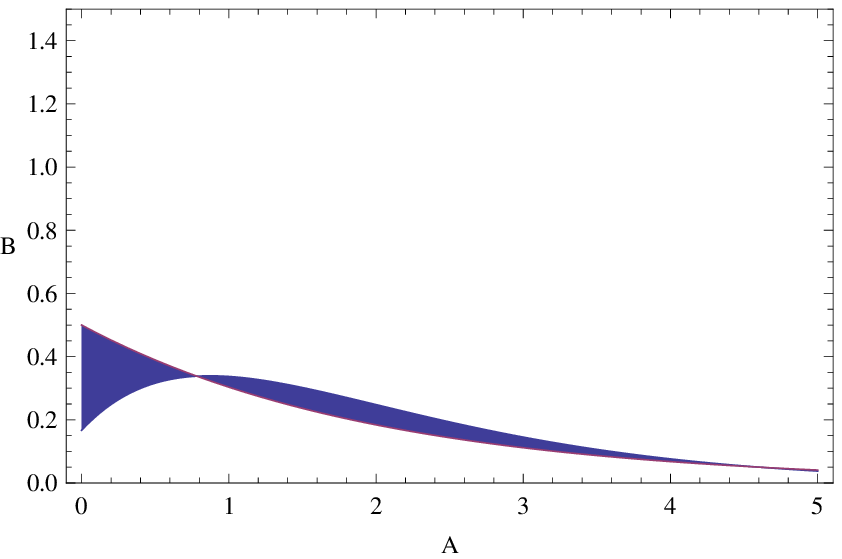} \hskip 2 mm
\psfrag{B}{} \psfrag{A}{\large\scriptsize (c)}
\includegraphics[width=1.5in, height=1.3in]{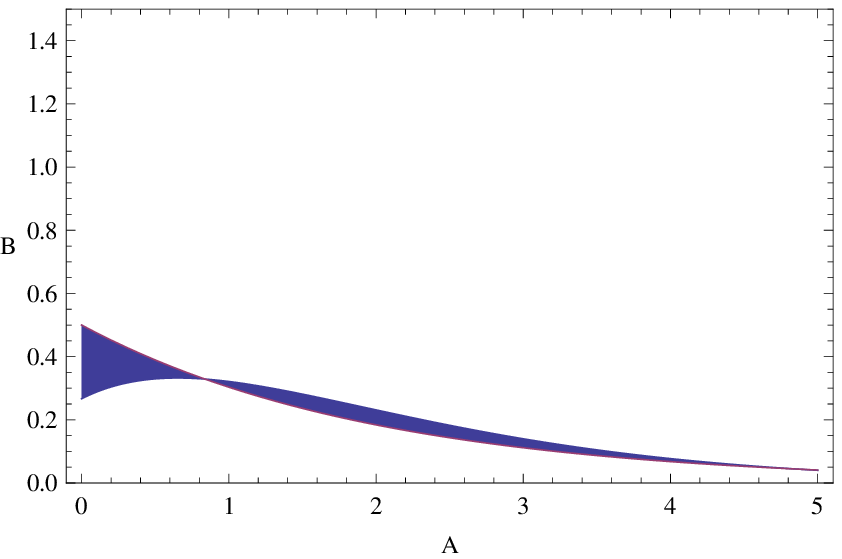}
\caption{Plot of $y(x)=xe^{-x}$, $T_{\lambda}$-iterates of $y$
for $\lambda=0.5$ and $\mu(x)={1\over 2}e^{-{1\over 2}x}$. (a)
$\vert\vert y-\mu\vert\vert$, (b) $\vert\vert
T_{\lambda}y-\mu\vert\vert$, (c) $\vert\vert
T_{\lambda}^2y-\mu\vert\vert$.} \label{fig-y-gamma1}
\end{center}
\end{figure}

\begin{example}
Take again the Gamma distribution $y(x)=xe^{-x}$, so that $y\in
B$ and $\delta={1\over 2}$, then in this case $\mu(x)={1\over
2}e^{-{1\over 2}x}$. For $\lambda= 0.5$, we find numerically that
$\vert\vert y-\mu\vert\vert=0.368226$, $\vert\vert
T_{\lambda}y-\mu\vert\vert=0.273011$, $\vert\vert
T_{\lambda}^2y-\mu\vert\vert=0.206554$, $\vert\vert
T_{\lambda}^3y-\mu\vert\vert=0.158701$, and so on. It is shown in
Fig. \ref{fig-y-gamma1}. Then we can guess that
$\lim_{n\rightarrow\infty}\vert\vert
T_{\lambda}^ny-\mu\vert\vert=0$. Observe that in this case
$T_{\lambda}y$ can be a non-decreasing function.
\end{example}

\section{Conclusions}

In this work, a continuous economic model recently
introduced \cite{lopezruiz2011} has been generalized.
This model takes into account idealistic characteristics
of the markets, where agents interact by pairs and exchange their money in
a random way. Also, the model implements a parameter that gives an idea of the
effectiveness of the agents when trading between them. A perfect effectiveness means
the total availability of the agents wealth to be gamed in their transactions and
a null effectiveness gives rise to a frozen market where each agent keeps intact
his money. We have shown in a rigorous manner that it does not matter the degree
of effectiveness of the market for the final statistical result.
In all the cases, these random markets evolve toward their asymptotic equilibrium,
that is, the exponential wealth distribution.

\section*{Acknowledgments}

R.L.-R. and J.L.L. acknowledge financial support from  the Spanish research projects
DGICYT-FIS2009-13364-C02-01 and DGICYT Ref. MTM2010-21037, respectively.

\end{document}